\documentclass[pdflatex,sn-mathphys-num,iicol]{sn-jnl}

\usepackage{graphicx}%
\usepackage{multirow}%
\usepackage{amsmath,amssymb,amsfonts}%
\usepackage{amsthm}%
\usepackage{mathrsfs}%
\usepackage[title]{appendix}%
\usepackage{xcolor}%
\usepackage{textcomp}%
\usepackage{manyfoot}%
\usepackage{booktabs}%
\usepackage{algorithm}%
\usepackage{algorithmicx}%
\usepackage{algpseudocode}%
\usepackage{listings}%
\UseRawInputEncoding

\unnumbered

\begin{document}

\title[Direct spin imaging detector based on freestanding magnetic nanomembranes with electron optical amplification]{Direct spin imaging detector based on freestanding magnetic nanomembranes with electron optical amplification}


\author*[1,2,3]{\fnm{O.E.}\sur{Tereshchenko}}\email{teresh@isp.nsc.ru}

\author[1]{\fnm{V.V.} \sur{Bakin}}

\author[3]{\fnm{S.A.} \sur{Stepanov}}

\author[1,2,3]{\fnm{V.A.} \sur{Golyashov}}

\author[1]{\fnm{A.S.} \sur{Mikaeva}}

\author[1]{\fnm{D.A.} \sur{Kustov}}

\author[1,4]{\fnm{V.S.} \sur{Rusetsky}}

\author[1,3]{\fnm{S.A.} \sur{Rozhkov}}

\author[1,3]{\fnm{H.E.} \sur{Scheibler}}

\author[4]{\fnm{A.Yu.} \sur{Demin}}

\affil[1]{\orgname{Rzhanov Institute of Semiconductor Physics, Siberian Branch, Russian Academy of Sciences}, \orgaddress{\city{Novosibirsk}, \postcode{630090}, \country{Russia}}}

\affil[2]{\orgname{Synchrotron Radiation Facility SKIF, Boreskov Institute of
Catalysis, Siberian Branch, Russian Academy of Sciences}, \orgaddress{\city{Koltsovo}, \postcode{630559}, \country{Russia}}}

\affil[3]{\orgname{Novosibirsk State University}, \orgaddress{\city{Novosibirsk}, \postcode{630090}, \country{Russia}}}

\affil[4]{\orgname{CJSC ``EKRAN FEP"}, \orgaddress{\city{Novosibirsk}, \postcode{630060}, \country{Russia}}}

\abstract{An analog of the optical polarizer/analyzer for electrons, a spin filter based on freestanding ferromagnetic (FM) nanomembrane covering the entrance of the microchannel plate (MCP) was applied for efficient spin filtering and electron amplification in the 2D field of view. To study the spin dependent transmission, we constructed a spin-triode device (spintron), which consists of a compact proximity focused vacuum tube with the Na$_{2}$KSb spin-polarized electron source, the FM-MCP and phosphor screen placed to run parallel to each other. Here, we demonstrate the fabrication of FM nanomembranes consisting of a [Co/Pt] superlattice deposited on a freestanding 3\,nm SiO$_{2}$ layer with a total thickness of 10\,nm. The FM-MCP has $\sim$\,10$^{6}$ channels with a single-channel Sherman function \textit{S}\,=\,0.6 and a transmission of $\sim$\,1.5$\times$\,10$^{-3}$ in the low electron energy range. The FM-MCP-based device provides a compact optical method for measuring the spin polarization of free electron beams in the imaging mode and is well suited for photoemission spectroscopy and microscopy methods.}

\maketitle

Despite the century-long history of electron spin discovery \cite{1_Uhlenbeck_1925} and the first attempt by Davisson and Germer to detect the polarization of electrons diffracted on a nickel crystal in 1929 \cite{2_Davisson_1929, 3_Kuyatt_1975}, the spin detection of a free electron remains a difficult task and much less efficient than the detection of the electron as a charge. Towards the end of the last century, it became clear that harnessing electron spin was crucial for the development of next-generation energy-efficient and high-speed devices, which in turn sparked an interest in materials science to study the spin texture of solids, both magnetic and non-magnetic materials \cite{4_Teruya_2014,5_Hasan_2010,6_Bentmann_2021,7_Miyamoto_2012}. The desire to gain insight into the spin-resolved band structure has driven the development of detection systems that could record the emitted electron energy, momentum, and spin in an ``all-in-one" photoemission experiment \cite{8_Bühlmann_2020,9_Christian_2020,9_5_Zhang_2022}. Therefore, spin-resolved and angle-resolved photoemission spectroscopy (spin-resolved ARPES) is at the forefront of solid state physics research due to its ability to provide valuable information on the spin-polarized electronic properties of materials. 

Detection of electron spin with the efficiency of spin-integrated angular-resolved photoelectron spectroscopy (ARPES) is considered to be a major issue in modern photoelectron spectroscopy, which initiated the search for an ``ideal" spin filter for free electrons\cite{10_Tereshchenko_2021}.  The definition of an ``ideal" spin detector can be formulated in terms of the capability of high-efficiency of single-channel detection transformed into multichannel (image-type) spin detection simultaneously with normal ARPES or momentum microscopy measurements. To increase efficiency (maximize signal-to-noise ratio, minimize the accumulation time), it is necessary to maximize the value $F = S^{2} I / I_{0}$ - figure of merit (FoM), defining the polarimeters sensitivity, where $I$ and $I_{0}$ are the intensity of the transmitted (scattered) and incident electrons, respectively, and $S$ is the Sherman function, defined as the observed scattering asymmetry when the incident electron is 100\% polarized. In order to further improve the detection efficiency, multichannel spin detection is desirable by measuring more than one data point at the same time.

Spin detectors developed to date are based on the phenomena caused by spin-orbit interactions (SOIs) of the electron \cite{11_Kirschner_1979}, the spin-exchange interactions of ferromagnetic (FM) materials \cite{12_Celotta_1979,12_5_Okuda_2017} or combination of both interactions \cite{10_Tereshchenko_2021}. Currently, it appears that the general tendency in spin ARPES follows the method of increasing the dimension of existing single-channel detectors, where the electron beam emitted from the sample is first projected onto the scattering target and then reconstructed onto the 2D electron sensitive detector\cite{13_Escher_2023}. For this reason, all the disadvantages inherent in single-channel detectors persist to the same extent for spatial resolution (image type) detectors, enhanced by the complex electron optics of image transfer.

Creating an analog of an optical polarizer/analyzer for electrons --- a free-electron spin filter with spatial resolution in combination with electron amplification --- might be considered as a constructive idea. This approach to measuring the spin polarization of an electron beam is to use a ``real filter" to inject spin-polarized electrons into a magnetic film. Spin dependent electron transmission through FM ultrathin films was proposed as a high-efficiency spin filter \cite{14_Schönhense_1993,15_Oberli_1998,16_Lassailly_1994,17_Weber_2001}. It was shown that the transmitted current depends on the relative orientation of the incident spin polarization with respect to the FM layer magnetization. Moreover, spin filters can act both as polarimeters and as sources of polarized electrons. All early attempts to create freestanding FM layers faced the need to make the films quite thick ($\geq$\,20\,nm), resulting in very low transmittance (10$^{-5}$--10$^{-7}$) while the most difficult task was to avoid pinholes in the film \cite{16_Lassailly_1994}. In addition, the lateral size of the prepared films was small enough (0.5--1\,mm) to give rise to the idea of a spatially resolved experiment. 

In this work, we demonstrate that freestanding FM nanomembrane covering the entrance of a microchannel plate can be used for efficient spin filtering and amplification in 2D field of view. In our previous works \cite{10_Tereshchenko_2021,18_Rusetsky_2022} we developed vacuum spin-LED based on a flat vacuum photodiode composed of two effective NEA semiconductor electrodes \cite{19_Rodionov_2017,20_Tereshchenko_2017,21_Golyashov_2020,22_Rozhkov_2024}. To characterize the 2D spin-filter we transformed a vacuum spin-LED into a spin-triode (spintron, an analog of triode lamp) for spin-polarized electron emission/injection by combining a freestanding magnetic nanomembrane (an analog of control grid) and a phosphor screen, which allowed us to calibrate the spin detector by measuring spin polarization (asymmetry) of the free electrons emitted from the recently discovered new Na$_{2}$KSb spin-polarized electron source \cite{18_Rusetsky_2022} with spatial resolution.

\begin{figure*}
\includegraphics[width=1.0\linewidth]{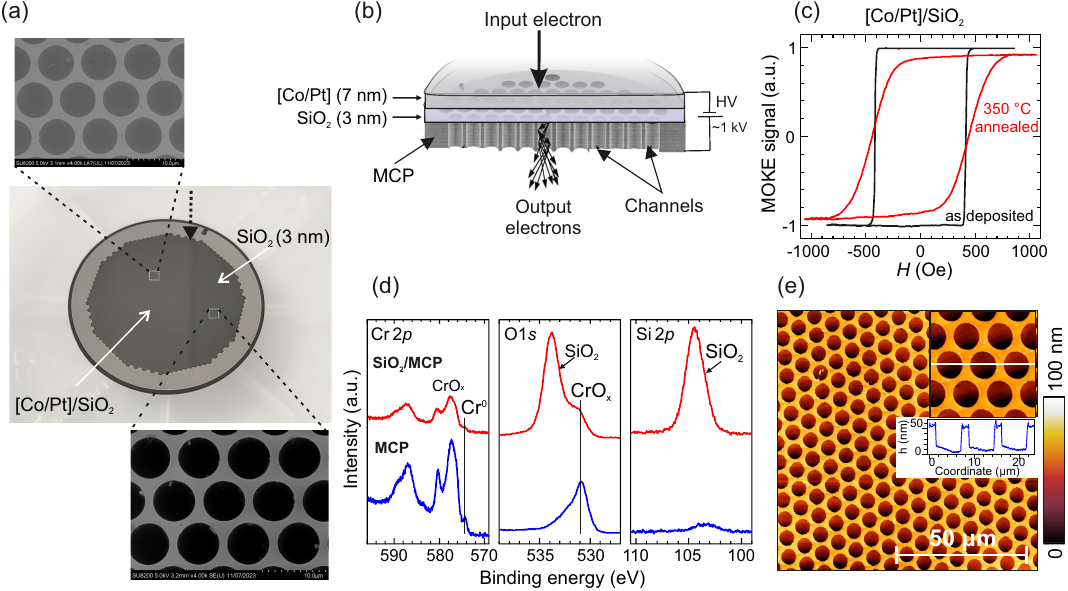}
\caption{\label{Fig.1} Formation and properties of freestanding [Co/Pt] multilayer on MCP. (a) Photograph of MCP with two differently covered parts, with the boundary shown by the arrow. A SiO$_{2}$ layer was first formed on the MCP, and then [Co/Pt] layers were deposited on the mask-free part of the MCP. Top-view SEM images of an array of glass microchannels coated with a $\sim 3$\,nm thick SiO$_{2}$ layer (bottom image) and coated with a [Co/Pt]/SiO$_{2}$ membrane (top image). (b) Operating principle of FM-MCP electron amplifier. When an incident spin-polarized electron passes through the FM nanomembrane and strikes a microchannel wall, secondary electrons are ejected. These electrons are accelerated through the microchannel, thereby creating an electron avalanche with a gain that depends on the number of stacked MCPs and is typically in the range of 10$^{3}$--10$^{7}$ (for one and two MCP). The electron cloud emanating from the bottom of the MCP is analyzed using a phosphor screen. (c) Magnetization hysteresis cycles for the out-of-plane magnetization in the [Co/Pt] superlattice grown on SiO$_{2}$ layer after deposition (black curve) and after subsequent anneal at 350\,$^{\circ}$C (red curve). (d) X-ray photoemission spectra of Cr\,2\textit{p}, O\,1\textit{s} and Si\,2\textit{p} core levels for the uncovered part of MCP surface (bottom row of spectra) and covered MCP by SiO$_{2}$ layer (top row). These measurements allowed us to estimate the thickness of the SiO$_{2}$ layer, which was found to be 3\,nm. (e) AFM image of [Co/Pt]/SiO$_{2}$ membrane covered MCP.}
\end{figure*}

\subsection{Formation of freestanding multilayer [Co/Pt]}\label{sec2}

The FM-MCP spin-filter comprises the [Co/Pt] superlattice Pt(2\,nm)/[Co(0.4\,nm)/Pt(1\,nm)]$_{3}$/Pt(1\,nm) about 7\,nm thick deposited on a 3\,nm SiO$_{2}$ layer covering the MCP with approximately 10$^{6}$ channels with an aperture single channel width of 6\,$\mu$m, shown in Fig.~\ref{Fig.1}(a),(b). The thickness of SiO$_{2}$ layer was determined from the X-ray photoemission spectra, shown in Fig.~\ref{Fig.1}(d). For this purpose, the XPS spectra were measured on the exposed surface of the MCP and on the one covered by the SiO$_{2}$ layer. From the attenuation of the Cr 2\textit{p} intensity line, the thickness of the SiO$_{2}$ layer was estimated to be 3\,nm. A similar procedure was applied to the FM film to determine the total FM thickness.

The magnetization $\boldsymbol{M}$ of the nanomembrane was monitored by the magneto-optical Kerr effect method. The out-of-plane hysteresis loop of as deposited [Co/Pt] structure is square and exhibits full magnetic remanence, as shown in Fig.~\ref{Fig.1}(c). No strong changes in the magnetic properties were observed after annealing FM-MCP at 350\,$^{\circ}$C, necessary for outgassing before spintron fabrication [Fig.~\ref{Fig.1}(c)]. Morphological studies of MCPs coated with [Co/Pt]/SiO$_{2}$ membrane by atomic force microscopy (AFM) showed that the membrane is deepened into MCP channels about 40\,nm [Fig.~\ref{Fig.1}(e)]. 

The spin selectivity was tested either by magnetizing the FM nanomembrane with an electromagnet or by changing the spin polarization of electrons by changing the angle of the quarter-wave plate excitation light of the photocathode. 

\begin{figure*}
\includegraphics[width=1.0\linewidth]{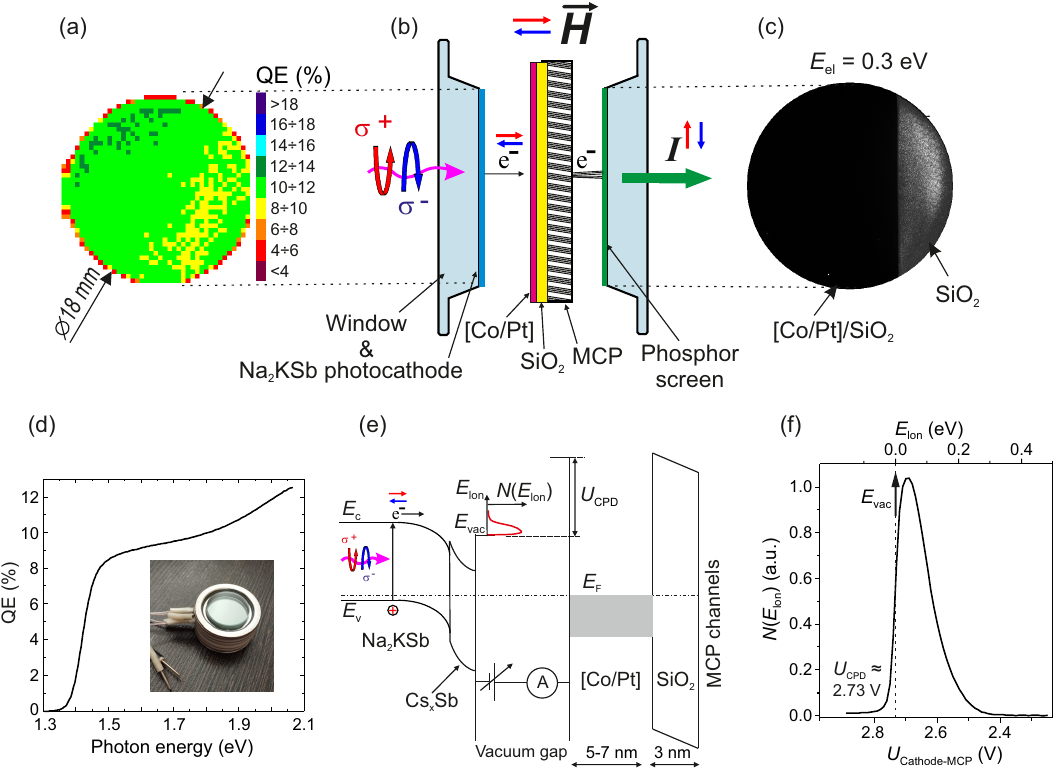}
\caption{\label{Fig.2} (a) Quantum efficiency map of the Na$_{2}$KSb photocathode measured under an excitation laser at 650\,nm (1.91\,eV). Each pixel represents one QE measurement; the legend on the right indicates the QE values in percent. The active area is 18\,mm in diameter. (b) Schematic presentation of the compact vacuum spin triode (spintron) for the investigation of the magnetic nanomembrane spin-selective properties. (c) MCP image for electrons with energy of 0.3\,eV injected into FM/SiO$_{2}$ nanomembrane, shown in Fig.~\ref{Fig.1}(a), under uniform photocathode illumination. The difference in intensity of FM/SiO$_{2}$ (dark) and SiO$_{2}$ (right light segment) is about two orders in magnitude. (d) Quantum efficiency of the photocathode as a function of photon energy (transmission mode). The vacuum spin-triode tube from the cathode side is shown in the inset. (e) The band diagram of Na$_{2}$KSb semiconductor electrode with the NEA and FM-MCP separated by a vacuum gap and electrically connected. (f) Energy distribution curve of photoemitted polarized electrons measured at photon excitation energies of 1.46\,eV in retarding-field FM-MCP electron detector. The position of the vacuum level $E_{\text{vac}}$ of the ferromagnetic film relative to the photocathode is 2.73\,eV, which corresponds to the contact potential difference $U_{\text{CPD}}$ between photocathode and FM-MCP.}
\end{figure*}

\subsection{Vacuum spin-triode (spintron): fabrication and properties}\label{sec3}

A schematic representation of a compact vacuum spin-triode device and a spin dependent injection study are shown in Fig.~\ref{Fig.2}(a-c). A photograph of the spintron is shown as an inset in Fig.~\ref{Fig.2}(d), which is constructed as follows: the Na$_{2}$KSb photocathode, FM-MCP spin-filter and phosphor screen were plane parallel mounted in an air-tight manner on the opposite flat sides of a cylindrical alumina ceramics body. A spin modulated electron beam is prepared using a Na$_{2}$KSb-type photocathode \cite{18_Rusetsky_2022}. Quantum efficiency spectrum $QE(\hbar\omega)$ of Na$_{2}$KSb photocathode measured in the transmission mode as a function of the incident photon energy is shown in Fig.~\ref{Fig.2}(d). The $QE$ is calculated as the ratio of the registered electrons per incident photons. Activation of the Na$_{2}$KSb surface by deposition of Cs and Sb leads to the formation of a negative effective electron affinity and an increase in quantum efficiency up to 0.2 electrons per incident photon \cite{22_Rozhkov_2024}. The quantum efficiency map of the Na$_{2}$KSb photocathode is shown in Fig.~\ref{Fig.2}(a) and demonstrates high $QE$ homogeneity. 

The spin dependent transmission experiment consists of injecting a longitudinally polarized electron beam photoemitted from a Na$_{2}$KSb source \cite{18_Rusetsky_2022} into an FM membrane with perpendicular magnetization. The electrons passing through the FM nanomembranes were amplified in the MCP channels ($\times 10^{3}$) and after leaving the MCP were projected onto the phosphor screen with further optical amplification ($\times$\,50--100). Fig.~\ref{Fig.2}(c) shows the image under homogeneous illumination of the photocathode. The difference in image intensity is due to the difference in the transmission coefficient of low-energy electrons ($E_{\text{el}} = 0.3$\,eV)  through the MCP covered by [Co/Pt]/SiO$_{2}$ (dark area) and SiO$_{2}$ (light area) and is equal to about 100 times. 

To study the electron emission and injection processes via the electron energy distribution, the photoelectron spectra are measured by differentiating the retarding voltage curves using the lock-in technique. This allows the longitudinal (along the beam) energy distributions of $N(E_{\text{lon}})$ photoelectrons to be measured using the spintron (photocathode --- FM-MCP gap) as a retarding field electron spectrometer \cite{19_Rodionov_2017}. The electron energy distribution curve (EDC) measured at room temperature for the Na$_{2}$KSb/Cs$_{x}$Sb photocathode with 1.38\,eV photon excitation energy is shown in Fig.~\ref{Fig.2}(f) and has the width (FWHM) of about 100\,meV. The low energy injection threshold into FM-MCP is given by the vacuum level of the [Co/Pt] entrance side [Fig.~\ref{Fig.2}(e)], whereas the lowest possible energy of the transmitted electron is determined by the barrier height at the [Co/Pt]/SiO$_{2}$ interface and/or the vacuum level $E_{\text{vac}}$ of the exit side of the SiO$_{2}$ layer. The electrons emerging from the [Co/Pt]/SiO$_{2}$ nanomembrane build up an energy distribution curve, which includes, besides the elastic electrons, a distribution of inelastically scattered electrons, which cannot be separated in FM-MCP structure. 

The asymmetry of the cathodoluminescence intensity (from the phosphor screen) was measured, which is proportional to the current difference between the electrons with spin up and spin down that passed through the spin filter at a fixed FM nanomembrane magnetization: $A(E) = (I(\boldsymbol{M})^{\text{up}} - I(\boldsymbol{M})^{\text{down}})/(I(\boldsymbol{M})^{\text{up}} + I(\boldsymbol{M})^{\text{down}})$, where $I(\boldsymbol{M})^{\text{up}}$ and $I(\boldsymbol{M})^{\text{down}}$ --- intensity of transmitted electrons with spin up and down at a fixed FM membrane magnetization. Another way to measure the asymmetry is to fix the electron polarization and measure the difference of currents passed through the spin filter at the opposite magnetization of the FM membrane: $A(E) = (I(+\boldsymbol{M})^{\text{spin}} - I(-\boldsymbol{M})^{\text{spin}})/(I(+\boldsymbol{M})^{\text{spin}} + I(-\boldsymbol{M})^{\text{spin}})$, where $I(+\boldsymbol{M})^{\text{spin}}$ and $I(-\boldsymbol{M})^{\text{spin}}$ --- intensity of transmitted electrons with a fixed electron polarization and opposite $\pm \boldsymbol{M}$ magnetization of the FM membrane. Magnetization of membranes results in a change in the electron attenuation length depending on the spin of the electrons with respect to the direction of magnetization: electrons with spin parallel to the magnetization will have a higher probability of transmission compared to electrons with antiparallel spin.

\subsection{Spin-filter characteristics}\label{sec4}

\begin{figure*}
\includegraphics[width=1.0\linewidth]{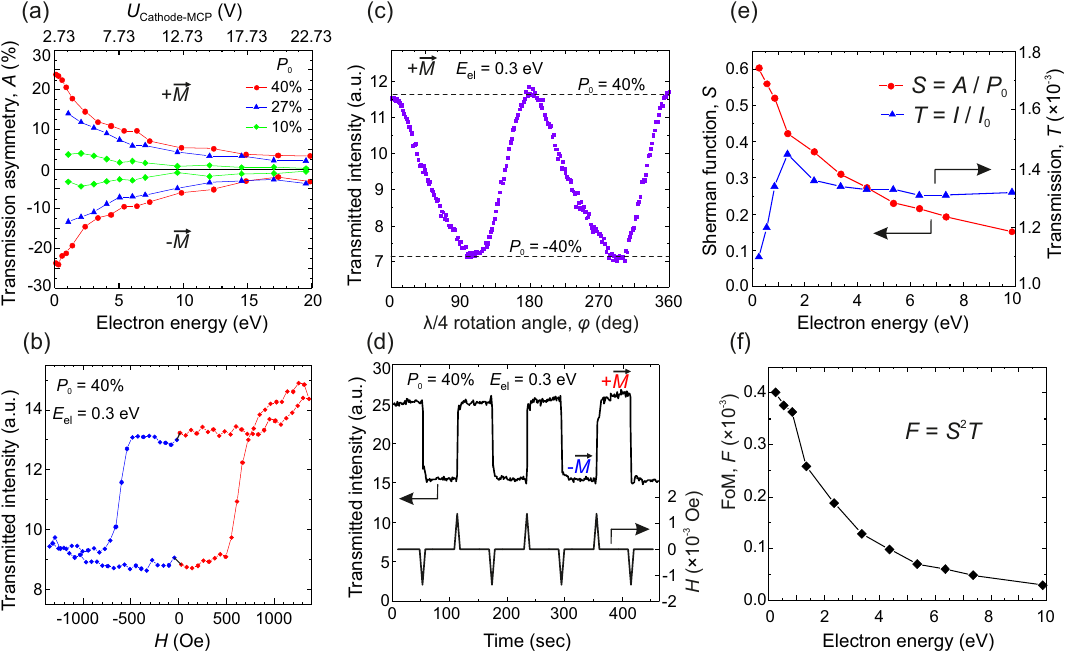}
\caption{\label{Fig.3} (a) Spin asymmetry $A(E) = (I(\boldsymbol{M})^{\text{up}} - I(\boldsymbol{M})^{\text{down}})/(I(\boldsymbol{M}^{\text{up}} + I(\boldsymbol{M})^{\text{down}})$ in \% as a function of the electron energy above $E_{\text{vac}}$ of [Co/Pt]/SiO$_{2}$ nanomembrane, measured for two opposite film magnetizations $\pm \boldsymbol{M}$ and various $P_{0}$ - polarization of the injected electron beam. The asymmetry is measured by inverting the direction of electrons spin parallel or antiparallel to the membrane magnetization. (b) Transmission current of spin polarized electrons through the FM-MCP as a function of applied external magnetic field with a magnetic field applied along the surface normal. (c) Dependence of the electron’s transmission on the quarter-wave plate rotation angle $\varphi$ at $E_{\text{el}} = 0.3$\,eV and fixed magnetization. The initial polarization of the electron beam passing through the FM nanomembrane $P_{0}$ changes its sign by 90-degree rotation of quarter-wave plate. (d) The transmitted spin-dependent current in a pulsed magnetic field $\pm \boldsymbol{H}$ that change the magnetization $\boldsymbol{M}$ of the [Co/Pt] membrane. (e) Sherman function $S = A / P_{0}$ and electron transmission coefficient $T = I / I_{0}$ for the [Co/Pt]/SiO$_{2}$ layer vs the energy above the vacuum energy $E_{\text{vac}}$. Transmission coefficient was measured as a ratio of currents passed through FM-MCP and uncovered MCP. (f) FoM $F = S^{2}I/I_{0}$ as a function of injected electron energy.}
\end{figure*}

To test the properties of the spin-filter, we started by measuring the asymmetry of electron transmission through the FM nanomembrane as a function of their energy. Figure ~\ref{Fig.3}(a) shows asymmetry $A(E) = (I(\boldsymbol{M})^{\text{up}} - I(\boldsymbol{M})^{\text{down}})/(I(\boldsymbol{M})^{\text{up}} + I(\boldsymbol{M})^{\text{down}})$ in \% of transmitted spin-polarized electrons through the [Co/Pt]/SiO$_{2}$ nanomembrane as a function of injection energy for opposite FM magnetization $\pm \boldsymbol{M}$ and different initial electron beam polarization $P_{0}$. The electron beam polarization $P_{0}$ was varied by changing the photon excitation energy of the Na$_{2}$KSb photocathode [18]. The zero kinetic energy is depicted relative to the vacuum level of the [Co/Pt] film, which was determined from measurements of longitudinal (along the beam) photoelectron energy distributions $N(E_{\text{lon}})$ using the Na$_{2}$KSb--FM-MCP gap as a retarding field electron spectrometer (Fig.~\ref{Fig.2}(e) and (f)) \cite{19_Rodionov_2017,23_Terekhov_1994}.

The asymmetry $A$ decreases with increasing electron energy as shown in Fig.~\ref{Fig.3}(a) for a [Co/Pt]/SiO$_{2}$ nanomembrane of about 10\,nm thickness, which is due to the decreasing matrix element for spin dependent scattering into the \textit{d}-shell of Co \cite{13_Escher_2023} and dilution by the secondary electrons. The secondary electrons have no memory of the spin polarization of the incident primary electrons and thus their intensity is independent of the initial spin configuration. Consequently, the intensity contribution due to secondary electrons lead to a reduction of the transmission asymmetry. Moreover, in the FM-MCP device, the elastically transmitted electrons through the [Co/Pt]/SiO$_{2}$ nanomembrane cannot be separated from the inelastically scattered ones, and the transmitted current $I = I_{0} e^{- \sigma d}$, where $\sigma$ is the spin dependent absorption coefficient, cannot be directly used for quantification. Nevertheless, it has been shown that the low-energy electrons polarization during FM membrane transmission within the energy loss of a few electron-volts, remains quite high \cite{15_Oberli_1998}.

One way to check a spin filter operation is to measure the current of transmitted polarized electrons at a fixed electron energy as a function of the membrane magnetization by an external magnetic field. In such a measurement, hysteresis should be observed in the dependence of the passing current, which can indeed be seen in Fig.~\ref{Fig.3}(b). 

Another way to test the efficiency of the spin filter is to measure the current of transmitted polarized electrons at a fixed electron energy and membrane magnetization as a function of electron polarization. The polarization of the electrons was varied by changing the quarter-wave plate angle of the excitation light of photocathode. In such a measurement, the dependence of the transmitted current should be proportional to the square of the cosine, which is shown in Fig.~\ref{Fig.3}(c).

\begin{figure*}
\includegraphics[width=1.0\linewidth]{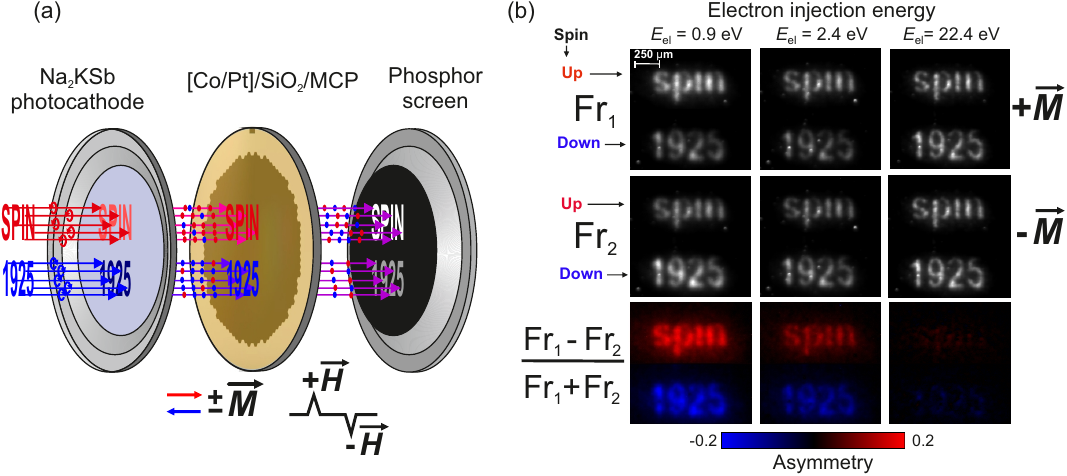}
\caption{\label{Fig.4} (a) Schematic representation of 2D pattern formation in polarized electrons: circularly polarized light is absorbed in the cathode and generates polarized electrons in the conduction band, which are emitted into vacuum, pass through a spin filter, are amplified in the MCP, accelerated, and fall onto a phosphor screen whose emission is recorded by a digital camera. (b) Images in spin-polarized electrons recorded with different injection energies and opposite magnetizations $\pm \boldsymbol{M}$ (without a permanent external magnetic field $\boldsymbol{H}$): frame Fr$_{1}$ measured after application of pulsed magnetic field $+\boldsymbol{H}$ and Fr$_{2}$ after application of opposite pulsed magnetic field $-\boldsymbol{H}$. The label ``spin" was created by excitation of spin-polarized electrons by light with right circular polarization, whereas the label ``1925" was created by photons with left circular polarization. The inscription commemorates the 100th anniversary of the discovery of electron spin.}
\end{figure*}

In order to determine the spin-detector efficiency FoM, it is necessary to acquire the Sherman function $S = A / P_{0}$ and transmission coefficient of injected electrons $T = I / I_{0}$. The Sherman function $S$ and transmission $T$ of the electron beam for the [Co/Pt]/SiO$_{2}$ nanomembrane are shown in Fig.~\ref{Fig.3}(e). To minimize the accumulation time (maximize signal-to-noise ratio), it is necessary to maximize the value $F = S^{2}I/I_{0}$, shown for the developed FM nanomembrane in Fig.~\ref{Fig.3}(f). The FM nanomembrane has a single-channel (spin-integrated) Sherman function $S \approx 0.6$ and a transmission of $1.4 \times 10^{-3}$ in the low electron energy range. This gives the value of the spin-integrated FoM equal to $0.4 \times 10^{-3}$. This value reflects the efficiency of the single-channel mode detector. The relatively large Sherman and transmission functions, the easy elimination of any instrumental asymmetry by magnetization reversal, and the possibility to operate in the electron counting mode make the FM-MCP very promising as a spin detector \cite{24_Drouhin_1996}. In order to further improve the detection efficiency, multichannel spin detection is desirable by measuring more than one data point at the same time \cite{10_Tereshchenko_2021}.

\subsection{Image-type spin resolved measurements}\label{sec5}

The development of a new spin detector with the capability of multichannel detection is in great demand for more efficient measurements in spin polarized low energy electron spectroscopy and microscopy. To illustrate the imaging capabilities of the FM-MCP spin filter, spin polarization images were recorded. The principle of image formation in spin-polarized electrons in a vacuum spin triode is shown in Fig.~\ref{Fig.4}(a). The image is formed by circularly polarized light, which is absorbed in the cathode and generates polarized conduction band electrons, which are emitted into the vacuum, pass through a spin filter, multiply in the MCP, then accelerate and fall onto a phosphor screen, whose emission is recorded by a digital camera. The two images (frames: Fr$_{1}$ and Fr$_{2}$) are measured at opposite magnetization of the FM membrane. Then, asymmetry $A = (\text{Fr}_{1} - \text{Fr}_{2}) / (\text{Fr}_{1} + \text{Fr}_{2})$ is plotted, as shown in Fig.~\ref{Fig.4}(b), for three different electron energies. The labels ``spin" and ``1925" are created by excitation of spin-polarized electrons by light with right and left circular polarization, respectively.

The spatial resolution of electron polarization can be seen quite well in Fig.~\ref{Fig.4}(b). In the case of two-dimensional polarization detection, we can introduce the efficiency of 2D spin-detector as $FoM_{\text{2D}} = FoM \times N_{\text{ch}}$, where $N_{\text{ch}}$ is the number of channels. The number of channels $N_{\text{ch}}$ is determined by dividing the detector area by the resolution of the experiment and is limited by the maximum number of channels in the MCP used, which is of the order of 10$^{6}$. This gives an estimate of $FoM_{\text{2D}}$ equal to 700. One advantage of 2D imaging, as opposed to integral detection \cite{17_Weber_2001}, is the ability to ignore the presence of pinholes. In images Fr$_{1}$ and Fr$_{2}$ in Fig.~\ref{Fig.4}(b), at least three pinholes can be seen as bright dots: one of them is located in the lower left corner, the second one in the middle of the digit ``1", and the third one under the letter ``n". It can be seen that the asymmetry intensity images are signal-free at the pinhole’s location, as they should be. At the current level of preparation of freestanding nanomembrane technology, we have reached a level where pinholes cover less than 1\% of the working detector area and can be easily accounted for in the digital image. 

Thus, we have demonstrated the ability of the FM-MCP detector to register an out-of-plane spin component. To measure in-plane spin components, it is necessary to grow an FM layer with an easy magnetization axis in-plane, which is much simpler than out-of-plane, and has already been used in previous spin-filtering experiments in the integral regime \cite{15_Oberli_1998}. In-plane two spin component detection experiments are being prepared in a real ARPES system \cite{25_Golyashov_2023} and will pave the way towards true vectorial imaging spin-detection.

\subsection{Conclusions}\label{sec6}

We presented a spin-triode device, characterization experiments, and test measurements with a two-dimensional direct imaging spin filter for the control and use of spin-polarized electrons and electron beams in a variety of spectroscopic and microscopic techniques \cite{26_Tusche_2024}. It is shown that the transmitted current depends on the relative orientation of the incident spin polarization with respect to the [Co/Pt] superlattice magnetization by measuring the direct transmission of a spin polarized free-electron beam through a freestanding [Co/Pt]/SiO$_{2}$ nanomembrane. The obtained results and developed technology will significantly improve the resolution of spin-resolved electron spectroscopy and microscopy and lead to wider applications of these techniques. The integration of such spin-filtering membranes in combination with a standard MCP in a hemispherical electron analyzer will provide massively parallel detection capabilities and one of the highest 2D efficiencies to date. Another important characteristic of the developed spin filter is its cost: it is far cheaper than commercially available spin filter for 2D imaging, including the single-channel Mott detector. The detector lifetime is determined by the stability of the MCP and is counted in years. Since the spin filter is an FM-coated MCP, it can be easily adapted to any modern electron spectrometer such as ARPES, momentum microscope, or even a spin-resolved LEED system. Freestanding magnetic layers, as shown here, may enable future vacuum spintronic devices for the development of a new field, vacuum spintronics.

\section{Methods}\label{sec7}

\subsection{FM-nanomembrane preparation and spintron fabrication.}\label{subsec1}

The [Co/Pt] superstructure is fabricated in an UHV metal deposition chamber on a substrate consisting of the 3\,nm thickness SiO$_{2}$ layer supported by an MCP with approximately 10$^{6}$ channels with an aperture width of 6\,$\mu$m. The SiO$_{2}$ is deposited on top of the nitrocellulose by evaporation of SiO$_{2}$ from a heated Mo crucible. After the dielectric was deposited, the MCP was annealed in air to burn out the nitrocellulose. On top of SiO$_{2}$ layer, [Co/Pt] layers are deposited by electron evaporation of Co and Pt from crucibles. Their thickness (as measured by a calibrated quartz microbalance) ranges from 0.4--2.0\,nm. The Co films are capped with a protecting Pt layer of 2\,nm thickness. Note that we used the direct evaporation method rather than the complex lift-off and transfer technique to form the freestanding layers \cite{27_Gu_2022}. The FM membrane is magnetized by pulsed magnetic field, allowing a choice of different magnetization states. 

The spin dependent experiments were carried out in a planar vacuum image intensifier type device consisting of the Na$_{2}$KSb photocathode on the glass substrate from one side, intermediate FM-MCP (spin-filter) and the phosphor screen from the other one, which are hermetically fixed parallel to each other at the opposite ends of an aluminum-oxide ceramic cylinder. The photocathode consists of an active polycrystalline Na$_{2}$KSb layer of 140\,nm thickness grown by vapor phase deposition on the glass of the input window and activated to NEA by Cs$_{x}$Sb layer \cite{18_Rusetsky_2022,22_Rozhkov_2024}. The diameters of the cathode and FM-MCP working area were 18\,mm, with the gap between the photocathode and FM-MCP of about 0.2\,mm.

\subsection{Film characterization }\label{subsec2}

X-ray photoelectron spectroscopy measurements were made using a SPECS GmbH ProvenX-ARPES system equipped with the ASTRAIOS 190 electron energy analyzer and the 2D-CMOS electron detector. XPS measurements were made using a focused monochromatic AlK$_{\alpha}$ radiation ($h\nu = 1486.7$\,eV, 0.2\,mm X-ray spot at 50\,W anode power). The surfaces of MCP samples were examined by scanning electron microscopy (LEO-1430) and atomic-force microscopy (NTEGRA PRIMA). The magnetization of the [Co/Pt] layers was monitored by the home-made magneto-optical Kerr effect system. The FM membranes are magnetized using electromagnet thus enabling the spin selectivity.

\subsection{Photoemission EDC measurements}\label{subsec3}

The energy distribution curves (EDC) for the emitted photoelectrons were measured by applying a potential between the photocathode and FM-MCP. The photoelectron spectra were measured by differentiating retarding voltage curves using the lock-in technique. The collector current is detected using the lock-in technique, slowly ramping the retarding voltage over the relevant electron energies and additionally modulating it with the amplitude of 5\,mV at the typical frequency of 140\,Hz. The resulting lock-in output signal is proportional to the electron beam energy distribution curve.

To produce spin-polarized electrons the Na$_{2}$KSb photocathode is illuminated by the 510\,nm (2.4\,eV), 635\,nm (1.95\,eV) and 850\,nm (1.46\,eV) laser diodes. The linear polarized laser beam is circularly polarized by a quarter-wave plate and falls on the backside of the photocathode through the photocathode window. The emitted electrons are longitudinally polarized along the normal to or against the detector surface. The photoemission current for measurements of transmission and asymmetry of FM membranes was in the range of 0.5--50\,nA. It is also possible to produce an unpolarized electron beam with $P_{0} = 0$ by using unpolarized light at the photocathode. The stability of the photoemission current from the photocathode was mainly determined by the stabilization of excitation laser intensity which was 0.1\%.



\begin{thebibliography}{29}
\ifx \bisbn   \undefined \def \bisbn  #1{ISBN #1}\fi
\ifx \binits  \undefined \def \binits#1{#1}\fi
\ifx \bauthor  \undefined \def \bauthor#1{#1}\fi
\ifx \batitle  \undefined \def \batitle#1{#1}\fi
\ifx \bjtitle  \undefined \def \bjtitle#1{#1}\fi
\ifx \bvolume  \undefined \def \bvolume#1{\textbf{#1}}\fi
\ifx \byear  \undefined \def \byear#1{#1}\fi
\ifx \bissue  \undefined \def \bissue#1{#1}\fi
\ifx \bfpage  \undefined \def \bfpage#1{#1}\fi
\ifx \blpage  \undefined \def \blpage #1{#1}\fi
\ifx \burl  \undefined \def \burl#1{\textsf{#1}}\fi
\ifx \doiurl  \undefined \def \doiurl#1{\url{https://doi.org/#1}}\fi
\ifx \betal  \undefined \def \betal{\textit{et al.}}\fi
\ifx \binstitute  \undefined \def \binstitute#1{#1}\fi
\ifx \binstitutionaled  \undefined \def \binstitutionaled#1{#1}\fi
\ifx \bctitle  \undefined \def \bctitle#1{#1}\fi
\ifx \beditor  \undefined \def \beditor#1{#1}\fi
\ifx \bpublisher  \undefined \def \bpublisher#1{#1}\fi
\ifx \bbtitle  \undefined \def \bbtitle#1{#1}\fi
\ifx \bedition  \undefined \def \bedition#1{#1}\fi
\ifx \bseriesno  \undefined \def \bseriesno#1{#1}\fi
\ifx \blocation  \undefined \def \blocation#1{#1}\fi
\ifx \bsertitle  \undefined \def \bsertitle#1{#1}\fi
\ifx \bsnm \undefined \def \bsnm#1{#1}\fi
\ifx \bsuffix \undefined \def \bsuffix#1{#1}\fi
\ifx \bparticle \undefined \def \bparticle#1{#1}\fi
\ifx \barticle \undefined \def \barticle#1{#1}\fi
\bibcommenthead
\ifx \bconfdate \undefined \def \bconfdate #1{#1}\fi
\ifx \botherref \undefined \def \botherref #1{#1}\fi
\ifx \url \undefined \def \url#1{\textsf{#1}}\fi
\ifx \bchapter \undefined \def \bchapter#1{#1}\fi
\ifx \bbook \undefined \def \bbook#1{#1}\fi
\ifx \bcomment \undefined \def \bcomment#1{#1}\fi
\ifx \oauthor \undefined \def \oauthor#1{#1}\fi
\ifx \citeauthoryear \undefined \def \citeauthoryear#1{#1}\fi
\ifx \endbibitem  \undefined \def \endbibitem {}\fi
\ifx \bconflocation  \undefined \def \bconflocation#1{#1}\fi
\ifx \arxivurl  \undefined \def \arxivurl#1{\textsf{#1}}\fi
\csname PreBibitemsHook\endcsname

\bibitem[\protect\citeauthoryear{Uhlenbeck and
  Goudsmit}{1925}]{1_Uhlenbeck_1925}
\begin{barticle}
\bauthor{\bsnm{Uhlenbeck}, \binits{G.E.}},
\bauthor{\bsnm{Goudsmit}, \binits{S.}}:
\batitle{Ersetzung der {H}ypothese vom unmechanischen {Z}wang durch eine
  {F}orderung bezüglich des inneren {V}erhaltens jedes einzelnen {E}lektrons}.
\bjtitle{Die Naturwissenschaften}
\bvolume{13},
\bfpage{953}--\blpage{954}
(\byear{1925})
\doiurl{10.1007/BF01558878}
\end{barticle}
\endbibitem

\bibitem[\protect\citeauthoryear{Davisson and Germer}{1929}]{2_Davisson_1929}
\begin{barticle}
\bauthor{\bsnm{Davisson}, \binits{C.J.}},
\bauthor{\bsnm{Germer}, \binits{L.H.}}:
\batitle{A {T}est for {P}olarization of {E}lectron {W}aves by {R}eflection}.
\bjtitle{Phys. Rev.}
\bvolume{33},
\bfpage{760}--\blpage{772}
(\byear{1929})
\doiurl{10.1103/PhysRev.33.760}
\end{barticle}
\endbibitem

\bibitem[\protect\citeauthoryear{Kuyatt}{1975}]{3_Kuyatt_1975}
\begin{barticle}
\bauthor{\bsnm{Kuyatt}, \binits{C.E.}}:
\batitle{Observation of polarized electrons by {D}avisson and {G}ermer}.
\bjtitle{Phys. Rev. B}
\bvolume{12},
\bfpage{4581}--\blpage{4583}
(\byear{1975})
\doiurl{10.1103/PhysRevB.12.4581} .
\bcomment{``It is found that results published by Davisson and Germer in 1929
  were analyzed incorrectly, and that they had in fact observed significant
  electron polarizations in the diffraction of low-energy electrons from
  single-crystal nickel."}
\end{barticle}
\endbibitem

\bibitem[\protect\citeauthoryear{Shinjo}{2014}]{4_Teruya_2014}
\begin{bchapter}
\bauthor{\bsnm{Shinjo}, \binits{T.}}:
\bctitle{1 - {O}verview}.
In: \beditor{\bsnm{Shinjo}, \binits{T.}} (ed.)
\bbtitle{Nanomagnetism and {S}pintronics ({S}econd {E}dition)},
pp. \bfpage{1}--\blpage{14}.
\bpublisher{Elsevier},
\blocation{Oxford}
(\byear{2014}).
\doiurl{10.1016/B978-0-444-63279-1.00001-0}
\end{bchapter}
\endbibitem

\bibitem[\protect\citeauthoryear{Hasan and Kane}{2010}]{5_Hasan_2010}
\begin{barticle}
\bauthor{\bsnm{Hasan}, \binits{M.Z.}},
\bauthor{\bsnm{Kane}, \binits{C.L.}}:
\batitle{Colloquium: {T}opological insulators}.
\bjtitle{Rev. Mod. Phys.}
\bvolume{82},
\bfpage{3045}--\blpage{3067}
(\byear{2010})
\doiurl{10.1103/RevModPhys.82.3045}
\end{barticle}
\endbibitem

\bibitem[\protect\citeauthoryear{Bentmann et~al.}{2021}]{6_Bentmann_2021}
\begin{barticle}
\bauthor{\bsnm{Bentmann}, \binits{H.}},
\bauthor{\bsnm{Maa\ss{}}, \binits{H.}},
\bauthor{\bsnm{Braun}, \binits{J.}},
\bauthor{\bsnm{Seibel}, \binits{C.}},
\bauthor{\bsnm{Kokh}, \binits{K.A.}},
\bauthor{\bsnm{Tereshchenko}, \binits{O.E.}},
\bauthor{\bsnm{Schreyeck}, \binits{S.}},
\bauthor{\bsnm{Brunner}, \binits{K.}},
\bauthor{\bsnm{Molenkamp}, \binits{L.W.}},
\bauthor{\bsnm{Miyamoto}, \binits{K.}},
\bauthor{\bsnm{Arita}, \binits{M.}},
\bauthor{\bsnm{Shimada}, \binits{K.}},
\bauthor{\bsnm{Okuda}, \binits{T.}},
\bauthor{\bsnm{Kirschner}, \binits{J.}},
\bauthor{\bsnm{Tusche}, \binits{C.}},
\bauthor{\bsnm{Ebert}, \binits{H.}},
\bauthor{\bsnm{Min\'ar}, \binits{J.}},
\bauthor{\bsnm{Reinert}, \binits{F.}}:
\batitle{Profiling spin and orbital texture of a topological insulator in full
  momentum space}.
\bjtitle{Phys. Rev. B}
\bvolume{103},
\bfpage{161107}
(\byear{2021})
\doiurl{10.1103/PhysRevB.103.L161107}
\end{barticle}
\endbibitem

\bibitem[\protect\citeauthoryear{Miyamoto et~al.}{2012}]{7_Miyamoto_2012}
\begin{barticle}
\bauthor{\bsnm{Miyamoto}, \binits{K.}},
\bauthor{\bsnm{Kimura}, \binits{A.}},
\bauthor{\bsnm{Okuda}, \binits{T.}},
\bauthor{\bsnm{Miyahara}, \binits{H.}},
\bauthor{\bsnm{Kuroda}, \binits{K.}},
\bauthor{\bsnm{Namatame}, \binits{H.}},
\bauthor{\bsnm{Taniguchi}, \binits{M.}},
\bauthor{\bsnm{Eremeev}, \binits{S.V.}},
\bauthor{\bsnm{Menshchikova}, \binits{T.V.}},
\bauthor{\bsnm{Chulkov}, \binits{E.V.}},
\bauthor{\bsnm{Kokh}, \binits{K.A.}},
\bauthor{\bsnm{Tereshchenko}, \binits{O.E.}}:
\batitle{Topological {S}urface {S}tates with {P}ersistent {H}igh {S}pin
  {P}olarization across the {D}irac {P}oint in {B}i$_{2}${T}e$_{2}${S}e and
  {B}i$_{2}${S}e$_{2}${T}e}.
\bjtitle{Phys. Rev. Lett.}
\bvolume{109},
\bfpage{166802}
(\byear{2012})
\doiurl{10.1103/PhysRevLett.109.166802}
\end{barticle}
\endbibitem

\bibitem[\protect\citeauthoryear{B\"{u}hlmann et~al.}{2020}]{8_Bühlmann_2020}
\begin{barticle}
\bauthor{\bsnm{B\"{u}hlmann}, \binits{K.}},
\bauthor{\bsnm{Gort}, \binits{R.}},
\bauthor{\bsnm{Fognini}, \binits{A.}},
\bauthor{\bsnm{D\"{a}ster}, \binits{S.}},
\bauthor{\bsnm{Holenstein}, \binits{S.}},
\bauthor{\bsnm{Hartmann}, \binits{N.}},
\bauthor{\bsnm{Zemp}, \binits{Y.}},
\bauthor{\bsnm{Salvatella}, \binits{G.}},
\bauthor{\bsnm{Michlmay}, \binits{T.U.}},
\bauthor{\bsnm{B\"{a}hler}, \binits{T.}},
\bauthor{\bsnm{Kutnyakhov}, \binits{D.}},
\bauthor{\bsnm{Medjanik}, \binits{K.}},
\bauthor{\bsnm{Sch\"{o}nhense}, \binits{G.}},
\bauthor{\bsnm{Vaterlaus}, \binits{A.}},
\bauthor{\bsnm{Acremann}, \binits{Y.}}:
\batitle{Compact setup for spin-, time-, and angle-resolved photoemission
  spectroscopy}.
\bjtitle{Rev. Sci. Instrum.}
\bvolume{91},
\bfpage{063001}
(\byear{2020})
\doiurl{10.1063/5.0004861}
\end{barticle}
\endbibitem

\bibitem[\protect\citeauthoryear{Tusche et~al.}{2020}]{9_Christian_2020}
\begin{barticle}
\bauthor{\bsnm{Tusche}, \binits{C.}},
\bauthor{\bsnm{Chen}, \binits{Y.-J.}},
\bauthor{\bsnm{Plucinski}, \binits{L.}},
\bauthor{\bsnm{Schneider}, \binits{C.M.}}:
\batitle{From photoemission microscopy to an ``{A}ll-in-{O}ne" {P}hotoemission
  {E}xperiment}.
\bjtitle{e-Journal of Surface Science and Nanotechnology}
\bvolume{18},
\bfpage{48}--\blpage{56}
(\byear{2020})
\doiurl{10.1380/ejssnt.2020.48}
\end{barticle}
\endbibitem

\bibitem[\protect\citeauthoryear{Zhang et~al.}{2022}]{9_5_Zhang_2022}
\begin{barticle}
\bauthor{\bsnm{Zhang}, \binits{H.}},
\bauthor{\bsnm{Pincelli}, \binits{T.}},
\bauthor{\bsnm{Jozwiak}, \binits{C.}},
\bauthor{\bsnm{Kondo}, \binits{T.}},
\bauthor{\bsnm{Ernstorfer}, \binits{R.}},
\bauthor{\bsnm{Sato}, \binits{T.}},
\bauthor{\bsnm{Zhou}, \binits{S.}}:
\batitle{Angle-resolved photoemission spectroscopy}.
\bjtitle{Nat. Rev. Methods Primers}
\bvolume{2},
\bfpage{54}
(\byear{2022})
\doiurl{10.1038/s43586-022-00133-7}
\end{barticle}
\endbibitem

\bibitem[\protect\citeauthoryear{Tereshchenko
  et~al.}{2021}]{10_Tereshchenko_2021}
\begin{barticle}
\bauthor{\bsnm{Tereshchenko}, \binits{O.E.}},
\bauthor{\bsnm{Golyashov}, \binits{V.A.}},
\bauthor{\bsnm{Rusetsky}, \binits{V.S.}},
\bauthor{\bsnm{Mironov}, \binits{A.V.}},
\bauthor{\bsnm{Demin}, \binits{A.Y.}},
\bauthor{\bsnm{Aksenov}, \binits{V.V.}}:
\batitle{A new imaging concept in spin polarimetry based on the spin-filter
  effect}.
\bjtitle{J. Synchrotron Rad.}
\bvolume{28},
\bfpage{864}--\blpage{875}
(\byear{2021})
\doiurl{10.1107/S1600577521002307}
\end{barticle}
\endbibitem

\bibitem[\protect\citeauthoryear{Kirschner and Feder}{1979}]{11_Kirschner_1979}
\begin{barticle}
\bauthor{\bsnm{Kirschner}, \binits{J.}},
\bauthor{\bsnm{Feder}, \binits{R.}}:
\batitle{Spin {P}olarization in {D}ouble {D}iffraction of {L}ow-{E}nergy
  {E}lectrons from {W}(001): {E}xperiment and {T}heory}.
\bjtitle{Phys. Rev. Lett.}
\bvolume{42},
\bfpage{1008}--\blpage{1011}
(\byear{1979})
\doiurl{10.1103/PhysRevLett.42.1008}
\end{barticle}
\endbibitem

\bibitem[\protect\citeauthoryear{Celotta et~al.}{1979}]{12_Celotta_1979}
\begin{barticle}
\bauthor{\bsnm{Celotta}, \binits{R.J.}},
\bauthor{\bsnm{Pierce}, \binits{D.T.}},
\bauthor{\bsnm{Wang}, \binits{G.-C.}},
\bauthor{\bsnm{Bader}, \binits{S.D.}},
\bauthor{\bsnm{Felcher}, \binits{G.P.}}:
\batitle{Surface {M}agnetization of {F}erromagnetic {N}i(110): {A} {P}olarized
  {L}ow-{E}nergy {E}lectron {D}iffraction {E}xperiment}.
\bjtitle{Phys. Rev. Lett.}
\bvolume{43},
\bfpage{728}--\blpage{731}
(\byear{1979})
\doiurl{10.1103/PhysRevLett.43.728}
\end{barticle}
\endbibitem

\bibitem[\protect\citeauthoryear{Okuda}{2017}]{12_5_Okuda_2017}
\begin{barticle}
\bauthor{\bsnm{Okuda}, \binits{T.}}:
\batitle{Recent trends in spin-resolved photoelectron spectroscopy}.
\bjtitle{J. Phys. Condens. Matter}
\bvolume{29},
\bfpage{483001}
(\byear{2017})
\doiurl{10.1088/1361-648X/aa8f28}
\end{barticle}
\endbibitem

\bibitem[\protect\citeauthoryear{Escher et~al.}{2023}]{13_Escher_2023}
\begin{barticle}
\bauthor{\bsnm{Escher}, \binits{M.}},
\bauthor{\bsnm{Weber}, \binits{N.B.}},
\bauthor{\bsnm{K\"{u}hn}, \binits{T.-J.}},
\bauthor{\bsnm{Patt}, \binits{M.}}:
\batitle{2{D} imaging spin-filter for {N}ano{ESC}a based on {A}u/{I}r(001) or
  {F}e(001)-p(1x1){O}}.
\bjtitle{Ultramicroscopy}
\bvolume{253},
\bfpage{113814}
(\byear{2023})
\doiurl{10.1016/j.ultramic.2023.113814}
\end{barticle}
\endbibitem

\bibitem[\protect\citeauthoryear{Sch\"{o}nhense and
  Siegmann}{1993}]{14_Schönhense_1993}
\begin{barticle}
\bauthor{\bsnm{Sch\"{o}nhense}, \binits{G.}},
\bauthor{\bsnm{Siegmann}, \binits{H.C.}}:
\batitle{Transmission of electrons through ferromagnetic material and
  applications to detection of electron spin polarization}.
\bjtitle{Annalen Der Physik}
\bvolume{505},
\bfpage{465}--\blpage{474}
(\byear{1993})
\doiurl{10.1002/andp.19935050504}
\end{barticle}
\endbibitem

\bibitem[\protect\citeauthoryear{Oberli et~al.}{1998}]{15_Oberli_1998}
\begin{barticle}
\bauthor{\bsnm{Oberli}, \binits{D.}},
\bauthor{\bsnm{Burgermeister}, \binits{R.}},
\bauthor{\bsnm{Riesen}, \binits{S.}},
\bauthor{\bsnm{Weber}, \binits{W.}},
\bauthor{\bsnm{Siegmann}, \binits{H.C.}}:
\batitle{Total {S}cattering {C}ross {S}ection and {S}pin {M}otion of {L}ow
  {E}nergy {E}lectrons {P}assing through a {F}erromagnet}.
\bjtitle{Phys. Rev. Lett.}
\bvolume{81},
\bfpage{4228}--\blpage{4231}
(\byear{1998})
\doiurl{10.1103/PhysRevLett.81.4228}
\end{barticle}
\endbibitem

\bibitem[\protect\citeauthoryear{Lassailly et~al.}{1994}]{16_Lassailly_1994}
\begin{barticle}
\bauthor{\bsnm{Lassailly}, \binits{Y.}},
\bauthor{\bsnm{Drouhin}, \binits{H.-J.}},
\bauthor{\bsnm{Sluijs}, \binits{A.J.}},
\bauthor{\bsnm{Lampel}, \binits{G.}},
\bauthor{\bsnm{Marli\`ere}, \binits{C.}}:
\batitle{Spin-dependent transmission of low-energy electrons through ultrathin
  magnetic layers}.
\bjtitle{Phys. Rev. B}
\bvolume{50},
\bfpage{13054}--\blpage{13057}
(\byear{1994})
\doiurl{10.1103/PhysRevB.50.13054}
\end{barticle}
\endbibitem

\bibitem[\protect\citeauthoryear{Weber et~al.}{2001}]{17_Weber_2001}
\begin{barticle}
\bauthor{\bsnm{Weber}, \binits{W.}},
\bauthor{\bsnm{Riesen}, \binits{S.}},
\bauthor{\bsnm{Siegmann}, \binits{H.C.}}:
\batitle{Magnetization precession by hot spin injection}.
\bjtitle{Science}
\bvolume{291},
\bfpage{1015}--\blpage{1018}
(\byear{2001})
\doiurl{10.1126/science.1057430}
\end{barticle}
\endbibitem

\bibitem[\protect\citeauthoryear{Rusetsky et~al.}{2022}]{18_Rusetsky_2022}
\begin{barticle}
\bauthor{\bsnm{Rusetsky}, \binits{V.S.}},
\bauthor{\bsnm{Golyashov}, \binits{V.A.}},
\bauthor{\bsnm{Eremeev}, \binits{S.V.}},
\bauthor{\bsnm{Kustov}, \binits{D.A.}},
\bauthor{\bsnm{Rusinov}, \binits{I.P.}},
\bauthor{\bsnm{Shamirzaev}, \binits{T.S.}},
\bauthor{\bsnm{Mironov}, \binits{A.V.}},
\bauthor{\bsnm{Demin}, \binits{A.Y.}},
\bauthor{\bsnm{Tereshchenko}, \binits{O.E.}}:
\batitle{New {S}pin-{P}olarized {E}lectron {S}ource {B}ased on {A}lkali
  {A}ntimonide {P}hotocathode}.
\bjtitle{Phys. Rev. Lett.}
\bvolume{129},
\bfpage{166802}
(\byear{2022})
\doiurl{10.1103/PhysRevLett.129.166802}
\end{barticle}
\endbibitem

\bibitem[\protect\citeauthoryear{Rodionov et~al.}{2017}]{19_Rodionov_2017}
\begin{barticle}
\bauthor{\bsnm{Rodionov}, \binits{A.A.}},
\bauthor{\bsnm{Golyashov}, \binits{V.A.}},
\bauthor{\bsnm{Chistokhin}, \binits{I.B.}},
\bauthor{\bsnm{Jaroshevich}, \binits{A.S.}},
\bauthor{\bsnm{Derebezov}, \binits{I.A.}},
\bauthor{\bsnm{Haisler}, \binits{V.A.}},
\bauthor{\bsnm{Shamirzaev}, \binits{T.S.}},
\bauthor{\bsnm{Marakhovka}, \binits{I.I.}},
\bauthor{\bsnm{Kopotilov}, \binits{A.V.}},
\bauthor{\bsnm{Kislykh}, \binits{N.V.}},
\bauthor{\bsnm{Mironov}, \binits{A.V.}},
\bauthor{\bsnm{Aksenov}, \binits{V.V.}},
\bauthor{\bsnm{Tereshchenko}, \binits{O.E.}}:
\batitle{Photoemission and {I}njection {P}roperties of a {V}acuum {P}hotodiode
  with {T}wo negative-{E}lectron-{A}ffinity {S}emiconductor {E}lectrodes}.
\bjtitle{Phys. Rev. Appl.}
\bvolume{8},
\bfpage{034026}
(\byear{2017})
\doiurl{10.1103/PhysRevApplied.8.034026}
\end{barticle}
\endbibitem

\bibitem[\protect\citeauthoryear{Tereshchenko
  et~al.}{2017}]{20_Tereshchenko_2017}
\begin{barticle}
\bauthor{\bsnm{Tereshchenko}, \binits{O.E.}},
\bauthor{\bsnm{Golyashov}, \binits{V.A.}},
\bauthor{\bsnm{Rodionov}, \binits{A.A.}},
\bauthor{\bsnm{Chistokhin}, \binits{I.B.}},
\bauthor{\bsnm{Kislykh}, \binits{N.V.}},
\bauthor{\bsnm{Mironov}, \binits{A.V.}},
\bauthor{\bsnm{Aksenov}, \binits{V.V.}}:
\batitle{Solar energy converters based on multi-junction photoemission solar
  cells}.
\bjtitle{Sci. Rep.}
\bvolume{7},
\bfpage{16154}
(\byear{2017})
\doiurl{10.1038/s41598-017-16455-6}
\end{barticle}
\endbibitem

\bibitem[\protect\citeauthoryear{Golyashov et~al.}{2020}]{21_Golyashov_2020}
\begin{barticle}
\bauthor{\bsnm{Golyashov}, \binits{V.A.}},
\bauthor{\bsnm{Rusetsky}, \binits{T.S.} \bsuffix{V.~S.~Shamirzaev}},
\bauthor{\bsnm{Dmitriev}, \binits{D.V.}},
\bauthor{\bsnm{Kislykh}, \binits{N.V.}},
\bauthor{\bsnm{Mironov}, \binits{A.V.}},
\bauthor{\bsnm{Aksenov}, \binits{V.V.}},
\bauthor{\bsnm{Tereshchenko}, \binits{O.E.}}:
\batitle{Spectral detection of spin-polarized ultra low-energy electrons in
  semiconductor heterostructures}.
\bjtitle{Ultramicroscopy}
\bvolume{218},
\bfpage{113076}
(\byear{2020})
\doiurl{10.1016/j.ultramic.2020.113076}
\end{barticle}
\endbibitem

\bibitem[\protect\citeauthoryear{Rozhkov et~al.}{2024}]{22_Rozhkov_2024}
\begin{barticle}
\bauthor{\bsnm{Rozhkov}, \binits{S.A.}},
\bauthor{\bsnm{Bakin}, \binits{V.V.}},
\bauthor{\bsnm{Rusetsky}, \binits{V.S.}},
\bauthor{\bsnm{Kustov}, \binits{D.A.}},
\bauthor{\bsnm{Golyashov}, \binits{V.A.}},
\bauthor{\bsnm{Demin}, \binits{A.Y.}},
\bauthor{\bsnm{Scheibler}, \binits{H.E.}},
\bauthor{\bsnm{Alperovich}, \binits{V.L.}},
\bauthor{\bsnm{Tereshchenko}, \binits{O.E.}}:
\batitle{Na$_{2}${KS}b/{C}s$_{x}${S}b interface engineering for high-efficiency
  photocathodes}.
\bjtitle{Phys. Rev. Appl.}
\bvolume{22},
\bfpage{024008}
(\byear{2024})
\doiurl{10.1103/PhysRevApplied.22.024008}
\end{barticle}
\endbibitem

\bibitem[\protect\citeauthoryear{Terekhov and Orlov}{1994}]{23_Terekhov_1994}
\begin{barticle}
\bauthor{\bsnm{Terekhov}, \binits{A.S.}},
\bauthor{\bsnm{Orlov}, \binits{D.A.}}:
\batitle{Fine structure in the spectra of thermalized photoelectrons emitted
  from {G}a{A}s with a negative electron affinity}.
\bjtitle{JETP Lett.}
\bvolume{59},
\bfpage{864}--\blpage{868}
(\byear{1994})
\end{barticle}
\endbibitem

\bibitem[\protect\citeauthoryear{Drouhin et~al.}{1996}]{24_Drouhin_1996}
\begin{barticle}
\bauthor{\bsnm{Drouhin}, \binits{H.-J.}},
\bauthor{\bsnm{Sluijs}, \binits{A.J.}},
\bauthor{\bsnm{Lassailly}, \binits{Y.}},
\bauthor{\bsnm{Lampel}, \binits{G.}}:
\batitle{Spin-dependent transmission of free electrons through ultrathin cobalt
  layers}.
\bjtitle{J. Appl. Phys.}
\bvolume{79},
\bfpage{4734}--\blpage{4739}
(\byear{1996})
\doiurl{10.1063/1.361656}
\end{barticle}
\endbibitem

\bibitem[\protect\citeauthoryear{Golyashov et~al.}{2023}]{25_Golyashov_2023}
\begin{barticle}
\bauthor{\bsnm{Golyashov}, \binits{V.A.}},
\bauthor{\bsnm{Kokh}, \binits{K.A.}},
\bauthor{\bsnm{Tereshchenko}, \binits{O.E.}}:
\batitle{Transport properties of ({B}i,{S}b)$_{2}${T}e$_{3}$ topological
  insulator crystals with lateral p-n junction}.
\bjtitle{Phys. Rev. Mater.}
\bvolume{7},
\bfpage{124204}
(\byear{2023})
\doiurl{10.1103/PhysRevMaterials.7.124204}
\end{barticle}
\endbibitem

\bibitem[\protect\citeauthoryear{Tusche et~al.}{2024}]{26_Tusche_2024}
\begin{barticle}
\bauthor{\bsnm{Tusche}, \binits{C.}},
\bauthor{\bsnm{Chen}, \binits{Y.-J.}},
\bauthor{\bsnm{Schneider}, \binits{C.M.}}:
\batitle{Low-energy spin-polarized electrons: their role in surface physics}.
\bjtitle{Frontiers in Physics}
\bvolume{12},
\bfpage{113814}
(\byear{2024})
\doiurl{10.3389/fphy.2024.1349529}
\end{barticle}
\endbibitem

\bibitem[\protect\citeauthoryear{Gu et~al.}{2022}]{27_Gu_2022}
\begin{barticle}
\bauthor{\bsnm{Gu}, \binits{K.}},
\bauthor{\bsnm{Guan}, \binits{Y.}},
\bauthor{\bsnm{Hazdra}, \binits{B.K.}},
\bauthor{\bsnm{Deniz}, \binits{H.}},
\bauthor{\bsnm{Migliorini}, \binits{A.}},
\bauthor{\bsnm{Zhang}, \binits{W.}},
\bauthor{\bsnm{Parkin}, \binits{S.S.P.}}:
\batitle{Three-dimensional racetrack memory devices designed from freestanding
  magnetic heterostructures}.
\bjtitle{Nat. Nanotechnol.}
\bvolume{17},
\bfpage{1064}--\blpage{1071}
(\byear{2022})
\doiurl{10.1038/s41565-022-01213-1}
\end{barticle}
\endbibitem

\end{thebibliography}
\end{document}